\documentclass[lettersize,journal]{IEEEtran}
\usepackage{amsmath}
\usepackage{amssymb}
\usepackage{stmaryrd}
\usepackage{mathtools}
\usepackage{acronym}
\usepackage{color}
\usepackage{xcolor}
\usepackage{verbatim}
\usepackage{booktabs}
\usepackage{siunitx}
\usepackage{suffix}
\usepackage{xstring}
\usepackage{xparse}
\usepackage{expl3}
\usepackage{mathrsfs}
\usepackage{tabularx}
\usepackage{multirow}
\usepackage{makecell}
\usepackage{array}
\usepackage[utf8]{inputenc} 

\usepackage[hidelinks]{hyperref}
\usepackage{cleveref}

\usepackage{graphicx}
\usepackage{subcaption}
\usepackage{wrapfig}

\usepackage[T1]{fontenc}    
\usepackage{amsfonts}       
\usepackage{microtype}      

\usepackage[compress,numbers]{natbib}

\AtBeginEnvironment{quote}{\singlespacing\small}
\usepackage{dirtytalk}
\usepackage{lscape}

\usepackage[ruled, linesnumbered]{algorithm2e}

\usepackage{wrapfig}

\usepackage{tikz}
\usetikzlibrary{arrows,backgrounds,calc}
\usepackage{relsize}
\usepackage{float}
\usepackage{kantlipsum} 
\usepackage{lipsum}
\usepackage{stfloats}

\usepackage{pifont}
\newcommand{\cmark}{\ding{51}}
\newcommand{\xmark}{\ding{55}}
\newcommand{\idkmark}{$\sim$}

\newcommand{\etal}{\textit{et al}.}
\newcommand{\etc}{\textit{etc}.}
\newcommand{\ie}{\textit{i}.\textit{e}.}
\newcommand{\eg}{\textit{e}.\textit{g}.}

\usepackage{csquotes}
\newcommand{\pquote}[2]{\begin{displayquote}``\textit{#1}" - P{#2}\end{displayquote}}

\newcommand{\sysacro}{{C}o{R}{T}{A}}

\begin{document}

\title{Lessons in Cooperation: A Qualitative Analysis of Driver Sentiments towards Real-Time Advisory Systems from a Driving Simulator User Study}

\author{Aamir Hasan$^{\dag}$, Neeloy Chakraborty$^{\dag}$, Haonan Chen$^{\dag}$, Cathy Wu$^{\ddag}$, and Katherine Driggs-Campbell$^{\dag}$
\thanks{$^{\dag}$A. Hasan, N. Chakraborty, H. Chen, and K. Driggs-Campbell are with the Department of Electrical and Computer Engineering at the University of Illinois Urbana-Champaign. Emails: {\tt\small \{aamirh2, neeloyc2, haonan2, krdc\}@illnois.edu}}%
\thanks{$^{\ddag}$C. Wu is with the MIT Laboratory for Information \& Decision Systems (LIDS), the Department of Civil and Environmental Engineering (CEE), and the Institute for Data, Systems, \& Society (IDSS) at the Massachusetts Institute of Technology, Email: {\tt\small cathywu@mit.edu}}%
\thanks{Manuscript received \today; revised \today.}}

\markboth{}%
{Hasan \etal: Lessons in Cooperation}

\IEEEpubid{0000--0000/00\$00.00~\copyright~2021 IEEE}

\maketitle

\begin{abstract}
Real-time Advisory (RTA) systems, such as navigational and eco-driving assistants, are becoming increasingly ubiquitous in vehicles due to their benefits for users and society.
Until autonomous vehicles mature, such advisory systems will continue to expand their ability to cooperate with drivers, enabling safer and more eco-friendly driving practices while improving user experience.
However, the interactions between these systems and drivers have not been studied extensively.
To this end, we conduct a driving simulator study (N=16) to capture driver reactions to a Cooperative RTA system.
Through a case study with a congestion mitigation assistant, we qualitatively analyze the sentiments of drivers towards advisory systems and discuss driver preferences for various aspects of the interaction.
We comment on how the advice should be communicated, the effects of the advice on driver trust, and how drivers adapt to the system.
We present recommendations to inform the future design of Cooperative RTA systems.
\end{abstract}

\section{Introduction}
\label{sec:intro}

Drivers everywhere rely on Real-time Advisory (RTA) systems almost every day, popularly in the forms of navigational~\cite{bark2014personal, bolton2015investigation, nguyen2016efficient, obuhuma2018real} and eco-driving assistants~\cite{chada2023evaluation, chada2022learning, chen2022driving, birrell2014effect, lin2014eco, lena2017vehicle}.
With the improvements in intelligent driving systems, new RTA systems are continually proposed to aid in driving scenarios such as controlled takeovers~\cite{kraus2016human, hock2016elaborating, large2019longitudinal, pakdamanian2022enjoy} and safe decision making~\cite{tran2013left, yan2016development, trabelsi2023advice, kim2016look}.
These RTA systems are deployed inside vehicles through onboard systems developed by car manufacturers in the form of Advanced Driver Assistant Systems (ADAS) at Society of Automotive Engineers (SAE) level 2 autonomy~\cite{sae_level}, or through the use of smartphones~\cite{lee2008effects,mamun2017intelligent}.
With increasing demand for such RTA systems, Autonomous Vehicles (AV) pose as the natural solution to encapsulate issues that these systems address.
However, the market penetration of AVs is likely to be delayed due to safety risks~\cite{kalra2014driving, dixit2016autonomous} and economic factors~\cite{mazein2024autonomous,stacy2018self,metz2021costly}, particularly in developing countries with large driving populations~\cite{zali2022autonomous}.
Thus, the development of these systems for such underserved markets, usually through SAE level 2 systems, can have a significant impact on \emph{our} future.

\begin{figure}[tb]
  \centering
    \vspace{-10pt}
    \includegraphics[width=0.5\textwidth]{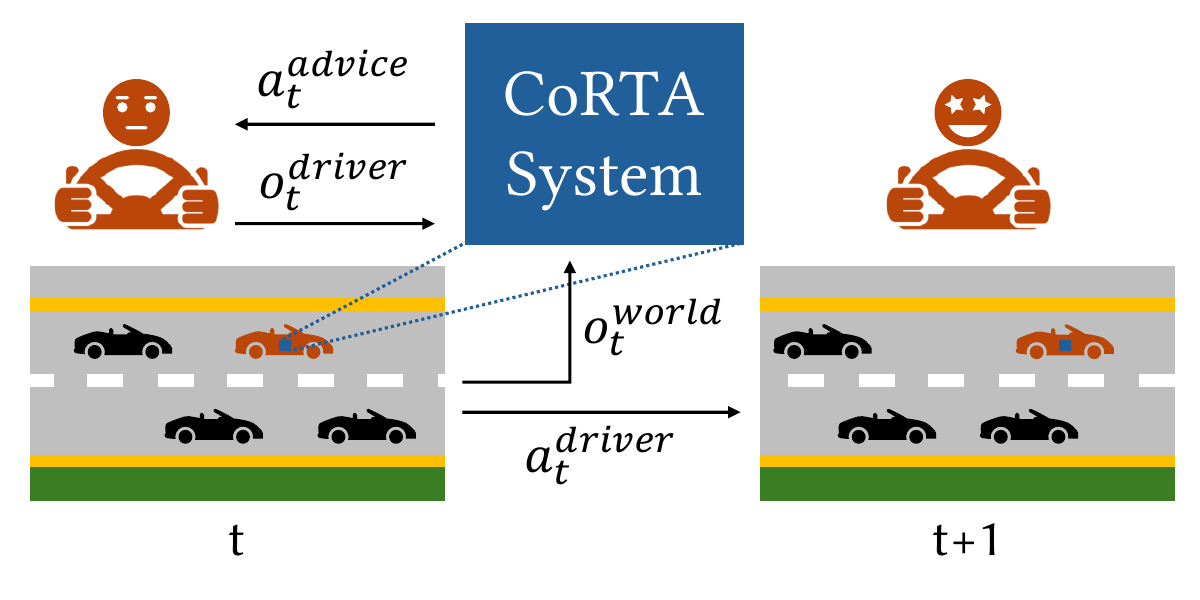}
    \caption{An illustration of real-time advisory systems helping drivers. The \sysacro~ system above observes drivers and their environment to provide advice to the driver and improve their user experience.}
    \label{fig:illustration}  
\end{figure}


These new developments continue to increase the time granularity of how frequently RTA systems advise drivers -- ranging from landmark-based events in navigation and turning assistants~\cite{nguyen2016efficient, obuhuma2018real} to second-by-second advice in eco-driving and speed advisory assistants~\cite{shahab2010auditory, jayawardana2022ecolearning, hasan2023perp}.
As this time granularity increases, the interactions between these systems and drivers must be formulated with careful consideration.
One avenue of improving user experiences with such time-precise systems is through the inclusion of cooperative features~\cite{walch2017from, pichen2021from, Woide2023Ive}, resulting in Cooperative Real-time Advisory (\sysacro) systems (illustrated in Figure \ref{fig:illustration}).
Such \sysacro~ systems are cognizant of driver behavior and preferences, and thus allow for enhanced interactions with drivers.
These interactions are likely to not only inform user experiences, but also to impact driver trust in automation, and hence, influence the efficiency of downstream Human-Vehicle teaming tasks (\eg~ eco-driving)~\cite{wintersberge2020explainable, walch2017from}. 
\IEEEpubidadjcol

The influence of AV behaviors on driver-vehicle interactions are widely studied at the SAE Level 3 or higher.
Particularly, the impact of different AV behaviors on driver attitudes (\eg~trust, comfort) towards these systems has been extensively researched, to inform future designs using driving simulator studies~\cite{Kraut2023AssertiveTR, Woide2023Ive, Helldin2013Presenting}, semi-structured interviews~\cite{Peng2023Exploring, wintersberge2020explainable}, and meta-surveys~\cite{Vongvit2022Meta, walch2017from}.
For example, a driving simulator study to capture the effects of assertive AV takeover maneuvers (\ie~the transfer of control of the vehicle between automation and the driver) by \citet{Kraut2023AssertiveTR} found that such behaviors invoked a faster response time in drivers without influencing driver stress or safety.
Similarly, ~\citet{Helldin2013Presenting} found an inverse effect of providing uncertainty information on driver trust in AVs during shared-control driving.

While such evaluations are abundant for systems with SAE Level 3 and above, similar treatments of driver attitudes for SAE Level 2 autonomy, particularly \sysacro~ systems, are sparse.
Works that do address these gaps are informative for future designs of \sysacro~ systems but recommend further studies on understanding how the system behaviors influence driver trust and reliance~\cite{deguzman2021knowledge}. 
Some works only evaluate singular factors such as the User Interface (UI)~\cite{kraus2016human, Wilfinger2010Influences}, while others primarily aim to test the effectiveness of the systems~\cite{birrell2014effect, hasan2023perp}.

Considering solely eco-driving \sysacro~ systems, previous works have used different evaluation approaches. 
Some works forgo human-in-the-loop testing and rely completely on software simulations, either due to limiting assumptions that render them hard to test, or unsatisfactory performance in non-idealized settings~\cite{asadi2011predictive, fleming2021incorporating, chen2022driving, hasan2023perp}.
Others such as \citet{birrell2014effect} test their solutions in naturalistic studies, but only discuss the efficiency of their methods and its impact on user behavior without evaluating driver preferences on the interactions.
Meanwhile, driving simulator studies that address driver preferences have been conducted but only evaluate certain aspects of the UI.
For example, \citet{shahab2010auditory} test auditory modes to communicate how drivers should change their speeds based on pitch and \citet{lena2017vehicle} test the function of a \emph{gamified} UI.
Both works successfully show the validity of their UIs, 
but do not discuss driver preferences for either the systems' behaviors or the advice provided by such systems.
On the other hand, \citet{chada2023evaluation} capture quantitative metrics on the potential impact and user acceptance of their novel system~\cite{chada2022learning}, but do not discuss qualitative opinions of drivers. 
Moreover, new learning-based eco-driving \sysacro~ systems that could significantly improve the efficiency of downstream tasks~\cite{kreidieh2018dissipating, vinitsky2018lagrangian, yan2021reinforcement, yan2022unified, han2024human, flow, pcp, hasan2023perp, cho2023temporal} are untested with humans.
Therefore, their interactions with users have not been studied extensively.
Thus, the following questions for interactions between drivers and (learned) \sysacro~ systems require more attention:
\begin{enumerate}
    \item[RQ1] What are driver expectations for the behavior of \sysacro~ systems? 
    \item[RQ2] How do Driver-\sysacro~ systems interactions affect driver behavior, trust in automation, and user experience?
    \item[RQ3] What are the UI requirements desired by drivers in their interactions with \sysacro~ systems?
\end{enumerate} 

In this article, we seek to address these questions in capturing driver sentiments towards \sysacro~ systems to enable better cooperation by combining a driving simulator study with semi-structured interviews.
Particularly, we first conduct a human-in-the-loop driving simulator study to allow participants to directly interact with \sysacro~ systems that exhibit a variety of behaviors.
Then, we conduct semi-structured interviews to qualitatively understand the impact of the diverse behaviors on drivers and their perceptions to capture user preferences for their interactions with \sysacro~ systems.
We analyze the sentiments expressed by drivers and discuss how their interactions can alter driver behavior.
Particularly, participants in our case study interacted with 9 \sysacro~congestion mitigation assistants in the form of eco-speed recommending Piecewise Constant Policies~\cite{pcp} (described in Section \ref{sec:method-policies}) and provided their firsthand sentiments on the interaction via semi-structured interviews. 
These recommendations and findings are relayed in this article to impact the design and development of future \sysacro~ systems.


Thus, our main contributions are:
\begin{enumerate}
    \item A driving simulator user study (N=16), focused on young adults, that captures their sentiments towards \sysacro~ systems through a case study on congestion mitigation assistants. 
    \item A qualitative analysis of driver preferences to provide insights into driver interactions with \sysacro~ systems. 
    \item Recommendations in designing future \sysacro~ assistants.
\end{enumerate}

To the best of our knowledge, we are the first to qualitatively investigate the broad sentiments held by drivers towards \sysacro~ systems through our user study. 

\section{Method}
\label{sec:method}
In this section, we first provide a short summary of the policies used in our \sysacro~ systems. 
We also introduce our user study, its procedure, and the user interface employed for the user study.

\subsection{The Advisory Policies}
\label{sec:method-policies}
As shown in Figure \ref{fig:illustration}, \sysacro~ systems are designed to provide a driver with some advice on how they should act in real-time while adapting to drivers' style and situational environmental factors.
Piecewise Constant (PC) Policies~\cite{pcp} exemplify such cooperative systems to address congestion mitigation, \ie~the reduction or dampening of stop-and-go waves in traffic~\cite{stern2018dissipation, flow}.
PC policies are policies trained using reinforcement learning that require the current action, \ie~the advice, to be held constant for a period of time -- referred to as the hold-length, $\delta$. 
\citet{pcp} designed this hold period to account for the time taken by drivers to react to and act upon the advice, and hence enabled the policies to be cooperative.
Residual variants~\cite{hasan2023perp} of such policies enable further cooperative behaviors through the use of driver trait inference modules. 
These policies aim to advise drivers of the speed that should be maintained in order to mitigate congestion.

In our case study, we use PC policies and two of their residual variants\footnotemark[1] that observe the driver and their environment to provide speed advice (See Appendix \ref{sec:appendix-policies} for details).
Particularly, we utilize 9 different policies: 3 policy types $\times$ 3 different hold-lengths: $\delta=\{5, 7, 10\}$ seconds.
The policies exhibited diverse behaviors in the speed advice they recommend and how the speed advice changes in real-time. 
For example, the advised speeds range between slow, ``just right'', or fast speeds, and there are either frequent or no changes, and even incremental or large changes in the advised speed. 
Some policies advise a sequence of speeds that when acted upon by drivers are intended to amount to a different, more effective speed advice \ie~such policies are not direct in communicating their advised speeds. 
Other policies are transparent and advise drivers on the most effective speed advice for their task.
While some systems may be more effective in mitigating congestion than others, all systems were cooperative and provided real-time assistance.
The analysis provided in this article does not focus on the individual differences in the policies, but on the effects of their diverse behaviors on their interactions with drivers.
For the remainder of this article, we use the phrases \emph{policy} and \emph{system} interchangeably. 
We also use the phrases \emph{advice} and \emph{speed} interchangeably.

\subsection{The Simulation Environment}
\label{sec:method-environment}
\begin{figure*}[t!]
    \centering
    \includegraphics[trim=0.cm 0cm 0cm 0cm, clip=True, width=\textwidth]{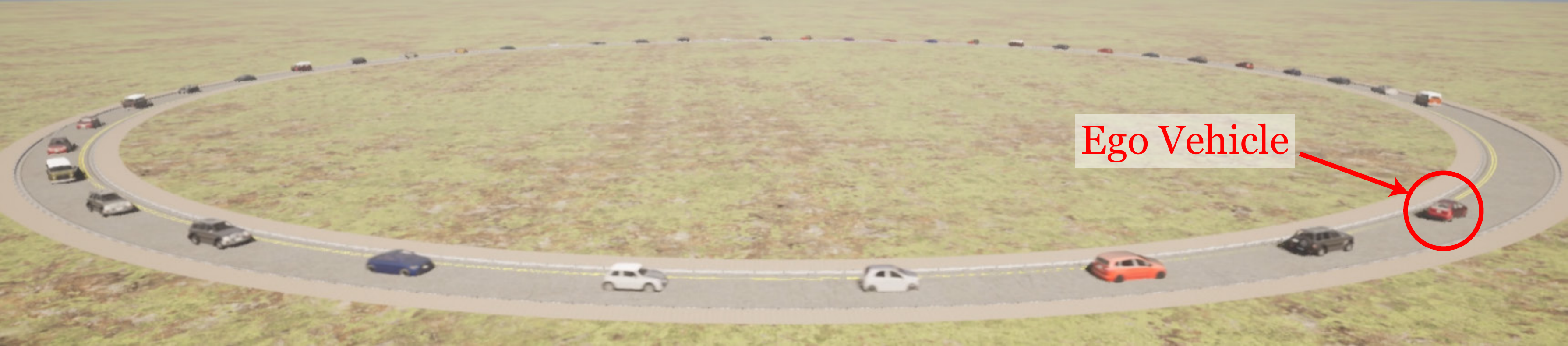}
    \caption{The simulation environment: A ring network with 40 vehicles, where one vehicle is controlled by the participant.}
    \label{fig:environment}
\end{figure*}

The simulated environment (shown in Figure \ref{fig:environment}) consists of $40$ vehicles driving on a single lane circular ring road with inner circumference of $628m$ and outer circumference of $654m$ in the CARLA Simulator~\cite{carla}.
Only one vehicle was controlled by the participant, while the other 39 vehicles were controlled using the IDM car following model~\cite{IDM}.
The weather in the simulation was set to a clear sunny day across all trials.
No other agents (other than the 40 vehicles) were present in the simulation to ensure minimal distractions to participants.
This simulation setup is designed to be consistent with the congestion mitigation literature~\cite{stern2018dissipation, flow, kreidieh2018dissipating}.

\begin{figure*}[b!]
    \begin{minipage}{0.35\textwidth}
        \centering
        \includegraphics[trim=0cm 0cm 0cm 0cm, clip=true, width=\textwidth]{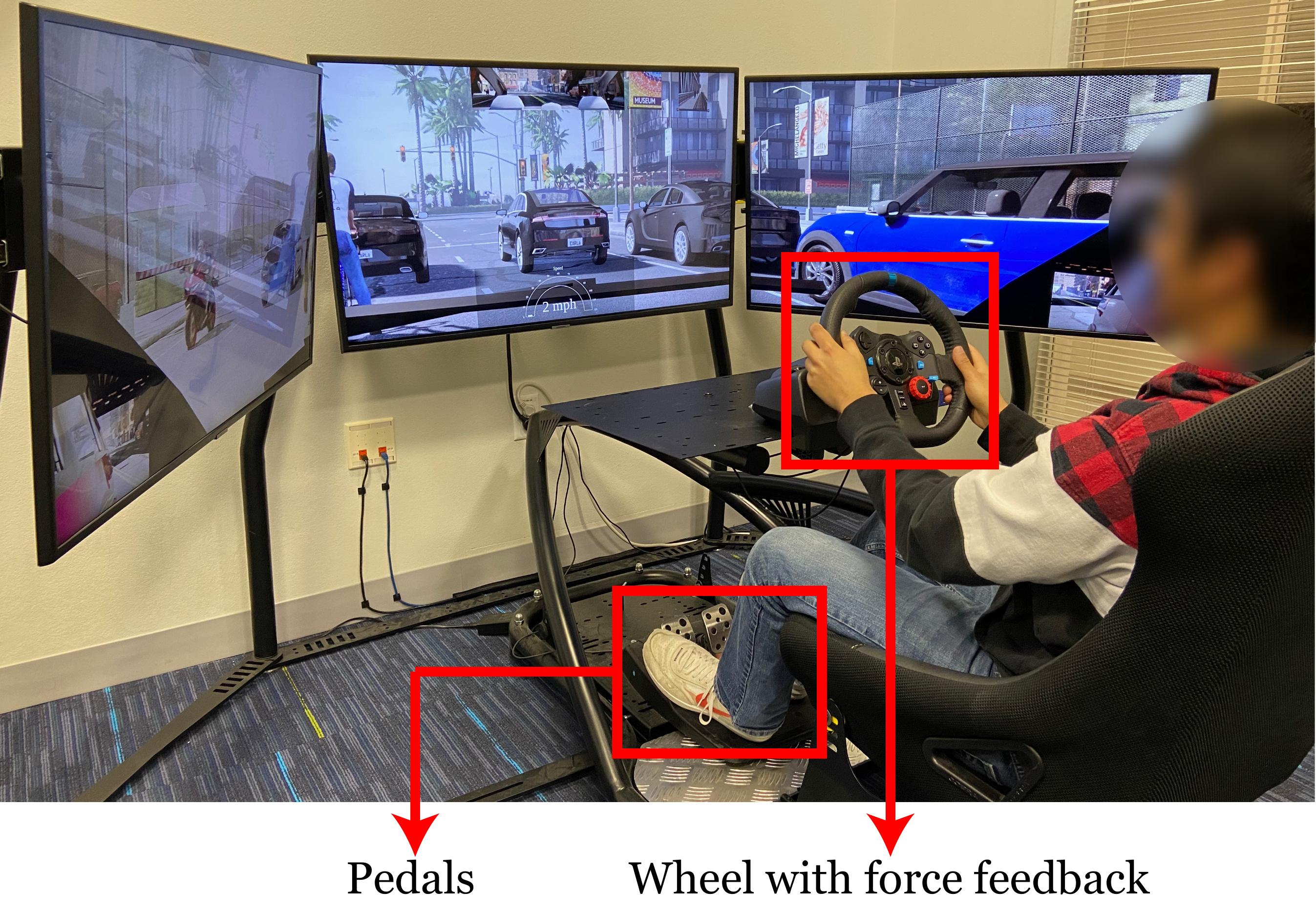}
        \caption{The driving simulator setup. 
        }
        \label{fig:driving-simulator}
    \end{minipage}\hfill
    \begin{minipage}{0.61\textwidth}
        \begin{subfigure}[t]{0.5\textwidth}
           \centering
           \includegraphics[trim=0.1cm 0cm 0cm 0.1cm, clip=true, width=\textwidth]{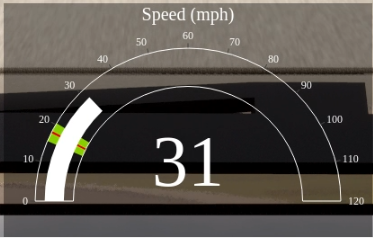}
            \caption{Driving outside the advised range.}
            \label{fig:ui-out} 
        \end{subfigure}
        \begin{subfigure}[t]{0.5\textwidth}
           \centering
            \includegraphics[width=\textwidth]{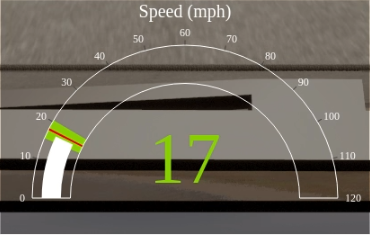}
            \caption{Driving inside the advised range.}
            \label{fig:ui-in}
        \end{subfigure}
        \caption{The user interface to convey the advice from the policy to the driver.}
        \label{fig:speedometer}    
    \end{minipage}
\end{figure*}

\subsection{The User Interface}
\label{sec:method-interface}
The participants' viewpoint is from the driver's seat, which includes the front view through the windshield and the rear-view mirror. 
The speedometer is shown at the bottom of the screen.
Figure \ref{fig:speedometer} shows the user interface for the study where the advice from the policies is communicated to the drivers via the speedometer, similar to other studies providing real-time speed advice~\cite{lin2014eco, Wilfinger2010Influences}.
The speedometer shows the speed of the vehicle in the form of a gauge and a number at the center of the gauge in a large font.
We display the advised speed on the speedometer as a red line surrounded by a $\pm1m/s$ ($\sim2.5$ mph) green range.
Participants were told that the red line indicated the exact speed that the policy was advising, but that they should try to keep their speed within the green range.
The green range effectively acted as an acceptable error bar for the drivers.
The number on the speedometer was highlighted in green when the participant's speed was within the acceptable error bar (See Figure \ref{fig:speedometer}(b)).


\subsection{The Procedure}
\label{sec:method-procedure}
\begin{figure*}[t!]
    \centering
    \includegraphics[width=\textwidth]{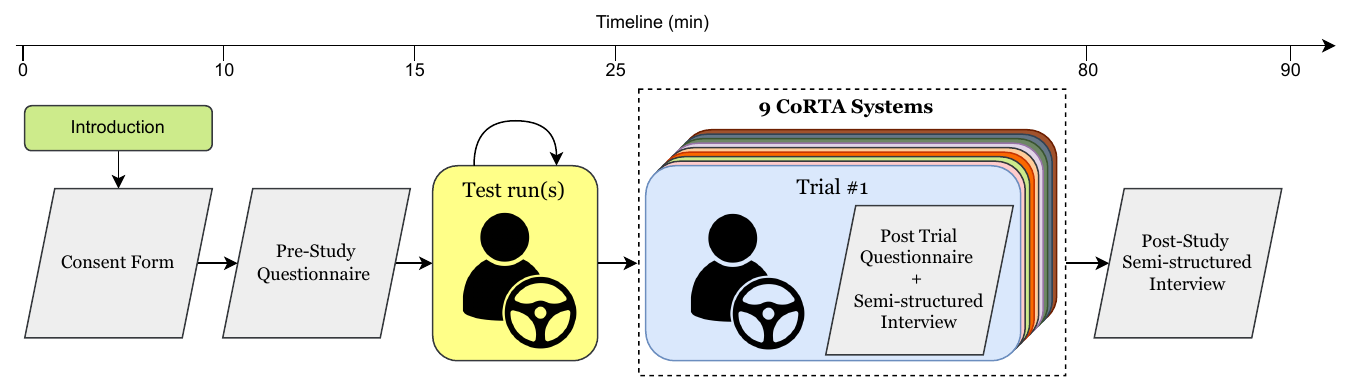}
    \caption{An overview of the user study procedure with the expected timeline for one participant.} 
    \label{fig:study-procedure}
\end{figure*}

Figure \ref{fig:study-procedure} illustrates an overview of the procedure for the user study. 
We conducted a 3-person pilot study before recruitment to validate the study procedure.
To begin the study, participants were first introduced to the goal of the study and briefed about their task. 
Participants were told that the aim of the study was to test congestion mitigation assistants and capture driver sentiments towards interactions with the assistants.
Additionally, participants were also told that their task was to (1) drive safely (avoid collisions, stay on-road, \etc) and (2) follow the advice provided by the system.
All participants were explicitly instructed to ignore the advice of the policy if they determined that it would lead to unsafe driving or a loss in control of the vehicle.
The study was conducted in a controlled lab environment and actions were taken to ensure that there were no external distractions present during the study procedures.

Participants first signed a consent form and then answered a pre-study questionnaire consisting of questions on demographics and driving experience. 
Participants were then given test runs to get familiar with using the driving simulator, the user interface, and interacting with the system.
Test runs were conducted in the same manner as the trials, until participants self-reported being comfortable controlling the vehicle\footnote{Only 2 participants required more than 2 test runs, all other participants only required 1 test run.}.
The policy used during the test runs was not evaluated during trials to prevent any biases. 
Subsequently, 9 different policies were tested in 5-minute trials, with 5-minute breaks every 3 trials.
The order of testing the 9 policies was uniquely randomized for each participant.
During each trial, participants first controlled their vehicle without any advice to enable the formation of stop-and-go waves.
After 90 seconds had elapsed, the policy began advising the participants for the next 200 seconds.
A trial was repeated if collisions occurred or if participants drove off the road\footnote{An average of 0.875 collisions and 0 off-roading related restarts occurred for each participant}.
After each trial, participants answered a post-trial questionnaire\footnotemark[2] that included the NASA Task Load Index (TLX) Questionnaire~\cite{hart1986nasa}, the System Usability Scale (SUS) Questionnaire~\cite{brooke1996quick}, and a question on how likely the participant was to use the policy in the real world. 
Following the post-trial questionnaire, a short semi-structured interview was conducted where the participants were queried about their experience during the trial.
At the end of all 9 trials, another semi-structured interview was conducted to gauge participants' views on the user interface and the systems as a whole.
In total, each participant was involved in the study for at most 90 minutes.
All interviews were audio recorded and transcripts of the recordings were analyzed after the end of the study.

\subsection{The Participants}
\label{sec:method-participants}
The IRB approved study was conducted in a between-subjects manner where all policies were tested by all participants in a unique, random order.
Our study population consisted of N=16 participants (10 male, 6 female) who were recruited using flyers and emails at the authors' university. 
No compensation was offered for participation in the user study.
We acknowledge that participants are likely to have been enthusiastic about driving simulators as a factor of our recruitment and compensation policy. 
Participants had a mean age of $23.7 \pm 2.8$ years where all were at minimum high school graduates or had earned their GED certificate or equivalent, with mean driving experience of $5.5 \pm 2.5$ years.
Participants were also informed not to discuss the study with anybody except researchers to avoid inadvertently inflicting biases on future participants. 


    
\section{Results and Discussion}
\label{sec:discussion}

\begin{table*}[t]
    \centering
    \caption{A summary of our recommendations for the design of future \sysacro~ systems, particularly real-time speed advisory systems. A more detailed table can be found in the Appendix as Table \ref{tab:recommedations-detailed}} 
    \label{tab:recommedations}
    \begin{tabular}{c c l c l}
        \toprule
        & Rec. & Behavior/attribute & Rec. & Behavior/attribute \\
        \midrule
        \multirow{6}{*}{Advice} 
        & \cmark
            &\emph{sweet spot}$^\dag$ speeds
            & \xmark
            & \makecell[l]{Extreme (too slow$^\dag$ or too fast$^\dag$) speeds} \\
        & \cmark
            & Smooth, gradual$^\dag$ changes in advice
            & \xmark
            & \makecell[l]{Large$^\dag$ change in magnitude} \\
        & \xmark
            & Frequent or sudden change in advice 
            & \xmark
            & No change in advice \\
         & \xmark
            & \makecell[l]{Oscillations between values in a set} 
            & \cmark
            & \makecell[l]{Transparent and direct} \\
        & \cmark
            & Matches user's driving style$^\dag$ 
            & \cmark
            & Reflects current traffic state \\
        & \cmark
            & Easy to understand 
            & \cmark
            & \makecell[l]{Adapts to driver's past history$^\dag$}\\
         \midrule
         \multirow{2}{*}{User Interface} 
         & \idkmark 
            & Speedometer 
            & \cmark 
            & \makecell[l]{Visible in peripheral vision}\\
            & \cmark 
            & \makecell[l]{Use of colors$^\dag$ to indicate changes}
            & \cmark 
            & \makecell[l]{Comfortable$^\dag$ Acceptable Error bar}
            \\
         \midrule
         \multicolumn{5}{c}{\cmark\quad Recommended \qquad\xmark\quad  Not Recommended \qquad\idkmark\quad  Controversial \qquad\dag\quad  Personalized} \\
        \bottomrule
    \end{tabular}
\end{table*}

In this section, we provide our conclusions on the \emph{do's} and \emph{don'ts} for \sysacro~ systems, that are summarized in Table \ref{tab:recommedations}.
All participant responses were edited by a research assistant to remove filler words (\eg~`like', `umm') to shorten responses that were verbose or repetitive.
Then, thematic coding was performed using NVivo14 on the responses by a single coder to seek sentiments relevant to our research questions. 
After all responses had been coded, resulting themes and significant quotes were identified and used to construct the analysis presented here.
To enhance quote understanding we replace verbiage and add context (between [brackets]) for some responses. 
We attest that the context and sentiments of participants are presented as they were intended.
Note that we refer to each participant as P\#\# where \#\# is the number assigned to them between 1 and 16.

{
}

\subsection{``What did you think of the advice?"}

\pquote{I think it's the best one... It seemed like the suggested speeds were very accurate, like if I did them, nothing bad would be happening.}{11}
Our interviews revealed multiple behaviors and characteristics of real-time advisory systems that significantly affected participant experiences. 
Unsurprisingly, the most significant attribute of the advice that drivers seek is accuracy (as it pertains to them).
Although in the quote above, P11 described accuracy as the ability of the policy to keep them safe, other factors such as the value of the advised speeds, the frequency of speed changes, the magnitude of the changes, time period to react to advice, and the understandability of the advice were all important contributors.
These behaviors affected drivers' trust in the system, and led to drivers adapting their driving style, and, in some extremes, ignoring the advice completely.


\textbf{The Value of the Advice:}
Users expect the advice provided by the system to be \emph{reasonable}, \ie~the speed should reflect the current state of the environment and task, and not request the users to take an action that can be seen as \emph{extreme} or out of line with social conventions.
Multiple participants describe the notion of a ``sweet spot" (P05) that they were hoping the advice would hit.
The specification for the \emph{sweet spot} varied between participants in terms of the advised speed range and the headway to their leader.
In our study, some participants preferred advice in the range of 25 to 30 mph, while others preferred advice in the slightly higher range of 30 to 35 mph.
Similarly, some drivers preferred leaving the length of one car's distance between them and their leader, while others preferred much larger distances of up to the length of three cars.
Systems that suggested advice corresponding to the driver's \emph{sweet spot} are viewed favorably.
\pquote{It actually kept me at... 25 mph for a longer period of time, which I liked. When the cars were like, you know, a [good] distance in front of me.
}{04}

In contrast, participants disliked when the advice did not fit their \emph{sweet spot} requirement. 
As it pertains to congestion mitigation, speed advice that was either too slow or too fast was not preferred and considered as ``extreme" (P04).
While drivers felt that speeds that were too slow were ``safe" (P13) and provided ``good control of the car" (P13), they also viewed them as counter-intuitive and mentioned that slow speeds led to inefficiencies, \ie~were ``causing a pile up" (P07) rather than help avoid one. 
Participants also reported feeling peer ``pressure of seeing that long line of cars behind [them]" (P10) when following such slow advice as it led to a violation of the social norms that they were accustomed to. 
Particularly, users were weary of being blamed by the cars behind them for holding up traffic when they were simply trying to follow the advice from the system. 
In addition to being stressed, frustrated, and ``irritated" (P10) by these aspects of slow advice, participants said that they would ``not want to follow these instructions, even if they [were] below the speed limit" (P07).
\pquote{It was [asking me to go] too slow. I feel like I was the one who was making this congestion, so I was a little stressed out when there was no cars in front [of me]... It was just frustrating.
}{02}

All 16 drivers reported that they felt very uncomfortable following advice that was too fast as it would result in unsafe behavior, such as collisions or off-roading.
Participants unanimously chose to ``just ignore the recommendation and follow [their] own mind[s]" (P12), on being recommended advice that was too fast.
Such advice was ``hard to follow" (P02) and made users ``lose control" (P12) of the vehicle and ultimately led to participants losing faith in the abilities of the system.
Thus, we recommend that \sysacro~ systems should carefully craft advice to appear accurate as defined by drivers and their personal \emph{sweet spots}.
We believe that this finding further motivates the development of adaptive technology in \sysacro~ systems that enables personalization to different drivers in line with the findings of \citet{walch2017from} for SAE Level 3 systems.

\textbf{The Changes in Advice:}
Participants disliked policies that advised significant speed jumps (\ie~large changes in the magnitude of the advice) and frequent or sudden unexpected changes.
The primary rationale for this sentiment was the increase in mental load in order to react to large and/or rapid changes that made driving less enjoyable. 
For example, P02 mentioned that the wavering of the recommended range between ``a high range and then suddenly going down to [a] little range... It was a little stressful for me."
Such behaviors also made the advice hard to follow and left drivers unsatisfied, ultimately leading to them ``not following it" (P03).

Additionally, advice that oscillated between a set of values, often between slow (\eg~15 mph) and fast (\eg~60 mph), was found to be undesirable and ``frustrating" (P04). 
P11 mentioned that such advice felt ``pointless," as trying to follow any of the speeds in the set was ``causing an issue" with the rest of the traffic, where they were forced between two undesirable behaviors -- they were either ``catching up to the rest of the pack or [they were] the one that was creating a pack". 
P04 also tried to rationalize this oscillatory behavior as an artifact of the system aiming to keep the driver engaged while encouraging them to maintain a certain speed.
They said that they wished the policy would be more \emph{transparent} and just advice the average of the set of speeds:
\pquote{It was constantly bouncing between 15/20 mph and 50/60 mph. And so, the effect was, realistically, my speed was somewhere between. Which I can see why it would want to keep me at that speed. But it was frustrating as a user... Well, if it wants me to go that speed, why doesn't it just put the bar there? Because that's where I'm driving, in effect, when it bounces back and forth like that.}{04}

Interestingly, on the other end of the spectrum, advice that was constant, \ie~did not change for a long period, was also not preferred as it led to driver disengagement, even though it was easy and (usually) safer to follow.
Drivers referred to policies suggesting such constant advice over multiple hold-length periods as ``conservative" (P02) (We discuss the effects of the hold-length itself in Section \ref{sec:results-hold-length}).
While constant advice was more acceptable than rapidly changing advice, it made drivers question if the policy was \emph{actually} cooperative as they felt like the system was ``less reactive" (P01). 
This constant behavior also led to interesting artifacts where the system would not ask drivers to slow down when they were approaching the car in front of them which users described as unsafe and dangerous.

\pquote{I think it would be better if the speed changed in a more continuous fashion and that's like feasible with the acceleration and deacceleration of the car. Instead of jumping from like 30 mph to 80 mph. If it slowly moved up that way, then it's okay -- I can follow that.}{04}

Drivers expressed that they preferred advice that was ``changing" because ``in this case it's a congestion. So, I think it makes sense to change the speed" (P12).
Similar to the \emph{sweet spot} in advice value, drivers prefer policies that provided smooth, gradual changes, where the rate of these changes is dependent on the drivers' style. 
Advice that changed gradually, \ie~in small increments, and allowed users to react comfortably enhanced the driving experience.
P11 in particular mentioned that having ``less big" jumps between consecutive actions made it ``more manageable to get to the" advised action.
They viewed such actions as ``something I could actually do" as opposed to ``an unreachable goal" that they attributed to the advice provided by the systems with large jumps.
Participants specifically mentioned that they would be more likely to use and follow a system that provided smooth advice to them.

Therefore, we further recommend that the cooperative systems should also include the drivers' preferred rate of change of advice in its \emph{sweet spot} determination.
Furthermore, we also recommend that \sysacro~ systems should be transparent and direct in their intent, \ie~the system should not communicate auxiliary advice that is meant to lead drivers into causing the effects of the actual intended advice.
Such manipulative interactions are viewed highly unfavorably by drivers.

\textbf{The Hold-length:}
\label{sec:results-hold-length}
The time that the advised action was designed to be held constant, \ie~the hold-length $\delta$ of the policies, had a significant effect on the users' experience. 
Generally, longer hold-lengths (10s in our case) are preferred as they allow users with ample time to follow the advice. 
Meanwhile, shorter hold-lengths (both 5s and 7s) either did not give enough time for drivers to react or were stress inducing.
Particularly, P13 mentioned that the policies with shorter hold-lengths were asking them to ``do too much in too little time."
Participants felt unsatisfied by policies that would change the advice as soon as they had achieved it as it left them frustrated and annoyed.
\pquote{The fact that it kept changing as soon as I reached the desired speed. The first question I marked here was this one [pointing to the frustration question. Researcher: ``Super frustrating?"] I think if it was a real-life scenario and it did that, I would just completely throw it out the window. [Talking to the policy] You don't deserve to be telling me what to do. You don't know what you're doing.}{10}

We hypothesize that there exists an ideal dynamic time-period that would satisfy drivers. 
Such an ideal time-period would depend on the drivers' personal preferences, attributes of the advice (\eg~the magnitude of change), and environmental factors.
A dynamic hold-length that depends on the above factors would ideally address the issues discussed above and lead to better, more reactive and cooperative systems.
We present quantitative results in the form of ANOVA tests on metrics collected from the questionnaires that corroborate these qualitative findings in Appendix \ref{sec:appendix-quant}.

\textbf{The Understandability of the Advice:}
Aside from wanting accurate advice, drivers are also particular about understanding the advice provided to them.
Systems that relayed advice that matched what drivers would do normally were touted as more ``realistic" (P14) and hence more usable.
P09 called their experience with a policy ``fun" because ``I like that it almost matched what we were doing... it almost made sense what kind of behavior it was trying to achieve."
\pquote{I did like it, because I think in that one it made sense was trying to be achieved and the driving speed was one that was more natural.}{09}

In contrast, advice that differed from the drivers' world view was seen as ``annoying" (P04) and ``frustrating" (P10), and hence influenced drivers to ignore or ``disobey" (P15) the system. 
For example, P02 called a system ``garbage" as its suggestions were ``the opposite of what I would do."
Additionally, advice that was perceived as confusing negatively impacted participants' trust in the systems' abilities.
\pquote{I was also confused why it was changing the speed on me when the cars in front of me were pretty far ahead for most of the simulation, so I didn't really understand why all of a sudden [it would] decide I should be going 5 miles[sic] faster, oh, now I need to go 5 miles[sic] slower. So, because of that, I was kind of like, okay, I don't know how much I should really follow this advice... I just, I couldn't understand why it was telling me what it was}{04}

In the case of traffic advisory systems, users prefer instructions that reflect the current state of traffic and allow them to blend into the flow of traffic, to be viewed as reasonable by others on the road.
For example, P05 mentioned that they ``didn't feel like [they were] fitting into like the rest of the traffic" when the policy was advising a speed of ``20 mph and there [were] no cars in front of [them]."
All drivers mentioned that they would be less likely to use systems that do not satisfy this requirement in real life and would rather ``throw [the system] out the window" (P10) and even reconsider purchasing vehicles with such systems.
\pquote{If I have to use a system then I will have to know why it's asked me to go that fast. Which is really unreasonable for me... So, I would definitely ditch this system. I would not buy the car.}{16}





We highly recommend that the advice given to drivers should also be easily understandable, in addition to the aforementioned personalizations. 
We foresee the inclusion of these personalizations as a major challenge for the development \sysacro~ systems as they would be required to address the various user needs \emph{and still} also fulfill the downstream tasks that they are designed to address.
Furthermore, participants noted their understanding of the system as a determinant of their trust in the system.


\pquote{It felt like it had no understanding of the distance between me and the car in front. So, because of that, I didn't really trust the system that much.}{04}

\textbf{Trusting the Advice:}
Users mentioned that the presence of inconsistencies during a trial decreased their trust in the system, even if they had experienced desirable behaviors during the trial. 
For example, P03 very bluntly stated during the post-study interview that ``I would not fully trust any of them", without providing a reason, even though they had previously expressed positive reactions to some systems.
We conjecture that this mistrust is due to a mismatch between their driving style and the policies' recommendations that was brought up during multiple trials: ``I'm in conflict with that advised speed a lot", ``mismatch is because of [my] driving style"
, ``I'm contradictory with the system." 

Furthermore, the contrast between the drivers' past driving experiences and the advice given by the system was a significant factor that influenced the drivers' trust in the system.
To this point, P10 recalled that the system ``was just going after every basic instinct and foundation of driving my father taught me." 
They specifically stated that this disagreement made them not want to listen to the advice: ``I don't like listening to this." 
P10 also added that as a consequence of this loss in trust they ``would listen to [the advice for] 2 minutes and then immediately ignore it" if they were driving ``in the real world" rather than in the simulator.

8 of 16 participants mentioned that they would ignore the advice in some form due to their loss in trust in the system.
Drivers wanted to partially ignore the advice if the system was somewhat reasonable and only faltered sometimes (\eg~provided incredibly high speed advice once while being stable otherwise). 
In particular, P16 noted that if the system ``gave me some like unreasonable commands that's pretty high, but also in some [cases] lower ones which are kind of reasonable," that they would ``just completely ignore the higher ones" and use those recommendations ``just like an indicator" to ``speed up a little bit".
In more extreme circumstances, \ie~if the system exhibited unstable behavior or inconsistencies, drivers reported wanting to completely ignore the system and were only following the advice solely because they were asked to for the study.
This is illustrated in the following interaction with P10:
\begin{displayquote}
\textbf{Researcher:} \textit{What did you think of this trial?}\\
\textbf{P10:} \textit{It was great once I opted to ignore all the inputs} [laughs]\\
\textbf{Researcher:} \textit{What did you think about what it was asking you to do?}\\
\textbf{P10:} \textit{Unreasonable. I think it was asking me to slow down when there was clearly a lot of room in front of me. Sometimes it was asking me to speed up when I could see [that] I would have to slam the brakes if I tried to do that. And so, I was like, well, I have a better decision making than this system.}  
\end{displayquote}

Therefore, we also recommend that \sysacro~ systems should adapt to the long-term driver behaviors and their past histories so that participants deem the system to be \emph{reasonable}. 
We posit that such adaption would lead to an increase in trust and therefore usage which would directly impact the downstream task that the system is designed to address.

{
}
\subsection{``What did you think of the User Interface?"}
Participants found the use of the speedometer to communicate the advice controversial.
P06 said that the UI was ``definitely pretty easy to follow... but at times it was a bit distracting." 
This sentiment was shared by 8 other participants, where P09, in particular, felt strongly that ``staring at the speedometer is not safe."
Users shared that driving was already a load incentive task due to the requirement of paying attention to the road and the other vehicles and adding ``any kind of instructions is actually another layer of burden that you impose on the driver" (P12). 
Thus, the increased load diverted drivers away from the main driving task ``overwhelmed" (P06) them and created opportunities for unsafe scenarios.
However, 3 of 9 participants noted that the use of the speedometer was a major issue only in the cases where the behavior of the advice increased the drivers' mental load.
Therefore, we advise selective use of this visual modality. 
\pquote{If it's not constantly changing, then I don't think it's an issue. But if it's constantly changing then I have to judge if the change is a valid one or not. And I have to constantly think about that, which is not good.}{12}

Some drivers mentioned that they viewed the advice solely as a suggestion and only kept the speedometer in their peripheral vision, as they would in real life.
These drivers mentioned that doing so ``eased the amount of mental load a little bit" (P05) but adversely affected how precise they were in following the advice.
Furthermore, the use of the green bar to indicate an acceptable range was touted by drivers as ``helpful" (P12) as it gave them more ``flexibility" (P14). 
P11 mentioned that the green bar facilitated easier instruction adherence as finding ``the right amount to press the pedal, for example, to get exactly on the red line is pretty difficult."
However, some drivers felt that ``the margin was a little small" (P16) and wished that the green range had been a little larger to allow for more leeway in how precisely the advice was followed, ``but it wasn't a big deal" (P16).


Participants also reacted positively to coloring the speed value (displayed on the speedometer) green when they were driving within the advised range and proclaimed it as a ``good touch" (P15).
P15 also mentioned that changing the color of the number ``made me look at the actual [gauge] less and just look at the big number and that didn't take as much attention".
The color choice also helped in minimizing the load on participants as they were ``looking for green" (P05) in their peripheral vision.
Therefore, we conclude that using attractive colors decreases the mental load on participants and enables easier instruction following.
Additionally, allowing a comfortable acceptable error bar (green range) would improve user experiences, but we find that this range differed between drivers.
We encourage further work to estimate the width of the acceptable error range based on the driver preferences.

The use of the color green and the dynamics of changing colors if the advice was followed, ``gamified (sic)" (P10) the system.
The green color scheme improved interactions with the system and also kept drivers engaged.
Participants were actively aiming to visualize green on their speedometer simply because ``seeing green is fun" (P10).
In this particular scenario, P10 was disappointed when the policy would change the advised action not soon after they achieved it because they did not ``get the satisfaction of getting the green number. And so that was frustrating." 
They went on to say that they would have been less frustrated ``if it gave me one little second of being green" (P10).
However, we note the induction of such ``gamified" reactions might be unsafe. 
We observed that it was easy for participants to get distracted and lose sight of ``what's in front of [them] as much because [they're] focused on putting the white[sic] inside the green part" (P09).

When asked about alternate suggestions for interface modes, one participant recommended that ``voice command[s] or the indicator light would be better" (P01) to interface with drivers.
While the 15 others preferred visual interfaces.
Heads-Up Displays (HUDs) were a popular alternative suggested by 3 participants (others had no opinions), as it would prevent them from looking down at the speedometer and hence allow them to focus on the road.
Some participants voiced a preference for being advised a change in action (\eg~go slower, maintain speed, go faster) over being advised an exact action.
Communicating this information through the use of directional arrows on the HUD was suggested.
We refrain from providing a staunch recommendation on the interaction mode as we did not quantitatively study the UI.



\subsection{``Did You Find Yourself Adapting to the System?"}
While users have preferences for the systems' behavior, we observed that drivers adapted their own driving during their interactions with the system in an effort to improve their experience.
Towards the latter trials, some users formed simple rules around when and how they would follow advice. 
These rules differed for each participant and were informed by their previous experiences with the different policies and their behaviors.
For example, P16 said that they would always check the distance to the car ahead of them first. 
Then if they determined the distance was large enough and their speed was less than or at 30 mph they would ``feel comfortable driving a little faster," if the policy suggested it. 
Else, they would only listen to the policy if it advised them to slow down. 
Otherwise, they would ``do nothing." 
Alternately, we observed that P07 would ignore all advice that was over 20 mph as they felt ``unable to control the vehicle at larger speeds", regardless of their distance to the car ahead of them.
While P07 did not provide a reason for their decision, we posit that they formed their rule due to their limited driving experience (1 year or less), and after almost crashing during an early trial.
Similarly, P15 mentioned that they also formulated similar rules and ``games" in real life:
\pquote{I do this when I'm driving a regular car as well, where I try to keep the speed at a certain number and just hold it there consistently. I think of it kind of like a game where you're holding the pedal down just enough where it's not going up or down.
}{15}

Additionally, participants also exhibited behaviors based on rationalizations that they arrived at by anticipating certain behaviors of the policy. 
When asked why P09 had driven faster than the advice given during a trial, they mentioned that ``it was going slow, and I didn't want it to go slow because the distance between the next car and us was quite large... We had just slowed down, so I knew we weren't going to slow down again any time soon."
Particularly, participants \emph{expected} the policies to behave similarly, even though the policies clearly exhibited different behaviors. 
Note that, information about the policies (if they were similar or different) was intentionally withheld from participants to observe if such expectations would arise. 
For example, in one instance P04 purposefully disregarded an advised action as they were anticipating the policy to change its advice at a later time, due to their experience in previous trials.
\pquote{I've been gaming the system a little bit because I've been expecting it to, you know, it'll recommend something high and then recommend something low. And this time it felt like it would recommend two things that were high. And then it'd be like, I guess I'm going to accelerate now. And then it would drop low, but it wouldn't drop it like far below what I had accelerated to. It felt like this is a little bit easier to match the speed. It's like I accelerated and now I just have to go back down a little bit.}{04}

Participants viewed the need to adapt to the system as normal and said that it was not cumbersome or annoying.
For example, P08 mentioned that using the system gave them ``the same feeling [they] get when [they] drive a car [they] haven't driven before".
P13 elaborated that using the system was akin to the ``standard experience" with ``any car you drive for the first time" where ``you kind of have to get used to the features."

\subsection{``Would you use these systems?"}
From the results of the questionnaires and interactions with participants we conclude that the policies tested during the study are not deployable without modifications, due to the drawbacks discussed previously.
However, while the systems might not suggest perfectly accurate advice, participants still found auxiliary uses for the policies. 
First, participants mentioned using the system to affirm their own driving styles and speeds.
Particularly, P12 mentioned the advised speed ``reassure[d] that the speed that [they] want to drive is correct" when the advice ``matches [their] optimal speed in [their] mind".
Others mentioned their interactions with the system made them feel ``very validated in [their] driving choices" (P10).
Second, users said that they would suggest the usage of this system as a teaching tool for new drivers. 
For example, P13 revealed: ``my younger sister is learning how to drive right now. If you told me I could put [the system] in her car, I would".
In P13's experience, they believed the \emph{conservative} systems that they experienced in some trials would aide in teaching their sibling how to drive safely.
Lastly, the use of the system in new environments was mentioned as a potential use case: ``If [I] was maybe driving in an unfamiliar environment, I'd try to follow this a bit more" (P14). 

Participants proclaimed that they would want to adopt \sysacro~ systems if there was a clear benefit for their usage.
P08 said that they would employ the system if they were ``convinced by whatever benefits" of the system. 
In particular, P08 mentioned environmental benefits of congestion mitigation agents, if proven to work, would motivate them.
Others agreed that they would use such a system were it be proven to have altruistic benefits (\eg~ environmental benefits). 
\pquote{So, I feel like it's kind of asking the user to [make a] contribution to society... It could result in a good way, in a sense that everyone contributes to that and make the whole[sic] traffic better. So, those type of [motivations], I buy that myself. So, I think it's good. I feel like, people would buy that. But there are definitely some users who just don't care.}{16}

In general participants had a positive outlook towards ideal \sysacro~ systems. P06 mentioned that this driving technology was ``something that should definitely be looked into for the future." They acknowledged that, while self-driving  solutions might address such goals, they thought that these systems are ``definitely something that should be looked into" for ``having advice for non-self-driving vehicles."
Therefore, we posit that future \sysacro~ systems built for altruistic purposes would be widely accepted by young drivers, \emph{if} they are personalized as postulated in our discussion.


\section{Conclusion and Future Work}
\label{sec:conclusion}

\textbf{Limitations and Future Work:}
While the findings presented above provide important insights into driver interactions with advisory systems, our methods are not without limitations. 
Firstly, we believe that a larger user study with a more diverse population is merited due to the smaller range in participant experiences in our study. 
Particularly, while our population addresses the sentiments of a significant portion of drivers, it does not reflect the needs of the majority.
Secondly, our participants experienced all study procedures in a lab environment with a driving simulator. 
While driving simulators 
are helpful in evaluating user sentiments 
in automotive settings, they lack the psychological and physical level feedback mechanisms offered by naturalistic driving. 
A naturalistic driving study in a controlled environment to evaluate \sysacro~ systems would provide further credence to our findings.
Lastly, our study was designed to capture qualitative user sentiments for one style of UI design. 
Performing a comparative study with multiple other UI options would be conducive in arriving at a definite UI framework for \sysacro~ systems.
Therefore, a systematic study that includes such definite quantitative methods to capture user perceptions would corroborate our findings. 

\textbf{Conclusions:} 
In this article, we present our findings from a  driving simulator user study (N=16) conducted with young adults.
We present insights into how drivers interact with cooperative real-time advisory systems and discuss various preferences and reactions for such interactions.
Our findings suggest that users prefer smooth-flowing, easy to understand advice that is also personalized to reflect their driving styles.
We also find that the use of attractive colors in the UI of these advisory systems improves user experiences and driver engagement. 
Additionally, agreement in systems' behavior and users' past experiences contributes significantly to drivers' trust in the system.
Lastly, find that drivers anticipate advice from the systems and adapt their driving mannerisms accordingly. 
We anticipate the use of these \emph{lessons in cooperation} in the design and development of future automotive assistive systems.

\section*{Acknowledgments}
The authors would like to thank all user study participants.
We would also like to thank Dr. Kristina Miller, Jung-Hoon Cho, Sirui Li, and Dr. Jeongyun Kim for their support and assistance with various aspects of this work.

\bibliographystyle{IEEEtranN}
\bibliography{root}

\newpage

{\appendix
\section{Quantitative Results based on the hold-length}
\label{sec:appendix-quant}
The quantitative results in Table \ref{tab:quantitative} are in agreement with the qualitative feedback discussed in Section \ref{sec:results-hold-length}.
The table shows aggregates for the Raw TLX Score computed from the NASA TLX questionnaire, the SUS Score, and the usage score computed with simple Likert scale statistical analysis (1=Very Unlikely; 5=Very Likely) on the responses to the question ``How likely are you to use this system while driving on the highway?".

\begin{table}[b!]
    \centering
    \caption{Quantitative Metrics from the user study}
    \begin{tabular}{c  c c c}
        \toprule
         Hold-length ($\delta$)
         & Raw TLX $(\downarrow)$ & SUS $(\uparrow)$ & Usage Score $(\uparrow)$ \\
         \midrule
         50 &  
         10.39 $\pm$ 3.76 &
         46.97 $\pm$ 22.41 & 
         1.85 $\pm$ 1.07 \\ 
         70 &
         8.81 $\pm$ 3.85 &
         59.43 $\pm$ 19.86 & 
         2.23 $\pm$ 1.09 \\ 
         100 & 
         \textbf{6.49 $\pm$ 2.57} &
         \textbf{69.89 $\pm$ 19.45} & 
         \textbf{2.81 $\pm$ 1.42} \\
         \bottomrule
    \end{tabular}
    \label{tab:quantitative}
\end{table}

We observe a clear trend where systems with larger hold-lengths are determined to be less load inducing (lower TLX) and more usable (higher SUS) in Table \ref{tab:quantitative}. 
Our findings are further corroborated by Univariate analysis of ANOVA tests that indicate statistically significant differences in the effect of the hold-length as the independent variable with the RAW TLX Score ($F_{2, 144} = 51.143, p < 0.01$) and the SUS Score ($F_{2, 144} = 40.456, p < 0.005$) as dependent variables, respectively.
While the TLX and SUS scores indicate that participants would prefer to use the systems with larger hold-lengths, the usage scores presented correspond to a general ``Neutral" feeling. 
Therefore, we conclude that the systems in their current state would not be desirable for use in the real world without enhancements to address driver concerns\footnotemark[1]. 
However, this low score is only a reflection of the current accuracy of the tested congestion mitigation systems, and not of the concept of \sysacro~ systems as a whole.
We envision a better reception for future real-time advisory systems that are developed to address the recommendations presented in this work.

We note here that neither the policy type nor the order in which policies were tested were statistically significant determinants for any qualitative metric. 
Particularly, for the policy type as the independent variable we observed $F_{2, 144}=4.179, p > 0.1$ and $F_{2, 144} = 0.504, p > 0.5$ with the RAW TLX and SUS Scores as dependent variables, respectively.
Additionally, we also observed no significant difference with the order in which the different policies were introduced to participants as the independent variable.
The ANOVA tests produced F statistics of $F_{8, 144}=0.280, p > 0.9$ and $F_{8, 144} = 0.228, p > 0.9$ for the Raw TLX and SUS scores as dependent variables, respectively.

 

\section{Piecewise Constant Policies}
\label{sec:appendix-policies}
Formally, PC policies, $\pi_{PC}: S \rightarrow A$, process an observed state $s_t \in S$ and output an action $a^{advice} = a_t \in A$, such that $a^{advice} = a_t = a_{t+1} = \dots = a_{t + \delta - 1}$. 
The driver is recommended the action $a^{advice}$ and acts upon it in the vehicle according to their wishes.
We refer interested readers to the works by \citet{pcp}, \citet{li2023stabilization}, \citet{cho2023temporal}, \citet{jayawardana2022ecolearning}, and \citet{hasan2023perp} for a more in depth discussion on PC policies and their variants\footnotemark[1]. 

In this work, we study cooperative real-time advisory systems through PC policies and two of its residual variants.
The three policies are:
\begin{enumerate}
    \item PC Policy (PCP)~\cite{pcp}: The PC policy introduced above.
    \item PC Residual Policy (RP)\footnotemark[1]: A modified PC policy that appends a residual action to the PCP action. Residual Policies trained with improvements in the reward function and the driver modeling to account for driver reactions to advice. 
    \item PC Personalized Residual Policy (PeRP)\footnotemark[1]: A modified RP that includes the use of a Variational autoencoder~\cite{vae} that performs unsupervised driver trait inference on how drivers react to advice.
\end{enumerate}

\begin{table*}[t]
    \centering
    \caption{A Details summary of our recommendations for the design of future \sysacro~ systems} 
    \label{tab:recommedations-detailed}
    \resizebox{\textwidth}{!}{
    \begin{tabular}{c c l l}
        \toprule
        & Rec. & Behavior/attribute & Why \\
        \midrule
        \multirow{11}{*}{Advice} 
         & \cmark
            &\emph{sweet spot}$^\dag$ speeds 
            & Easy to follow \\
         & \xmark
            & \makecell[l]{Extreme (too slow$^\dag$ or too fast$^\dag$) speeds} 
            & Leads to inefficiencies, Drivers feels peer pressure \\
         & \cmark
            & Smooth, gradual$^\dag$ changes in advice 
            & Easy to follow \\
         & \xmark
            & \makecell[l]{Large$^\dag$ change in magnitude} 
            & Hard to enact \\
         & \xmark
            & Frequent or sudden change in advice 
            & Hard to follow, Stressful, Leads to violation of social norms  \\
         & \xmark
            & No change in advice 
            & Perceived as non-reactive or absent \\
         & \xmark
            & \makecell[l]{Oscillations between values in a set} 
            & \makecell[l]{Manipulative, Stressful, Decreases trust in system} \\
         & \cmark
            & \makecell[l]{Transparent and direct} 
            & \makecell[l]{Increases trust in system, Does not confuse users} \\
         & \cmark
            & Matches user's driving style$^\dag$ 
            & \makecell[l]{Perceived as reasonable, Increases trust in the system} \\
         & \cmark
            & Reflects current traffic state 
            & \makecell[l]{Increases trust by allowing driver to blend into traffic}\\
        & \cmark
            & Easy to understand 
            & Increases trust, easy to follow \\
        & \cmark
            & \makecell[l]{Adapts to driver's past history$^\dag$}
            & \makecell[l]{Perceived as reactive, increases trust in system} \\
         \midrule
         \multirow{6}{*}{User Interface} 
         & \idkmark 
            & Speedometer 
            & \makecell[l]{Easy to understand, Distracting and hence dangerous} \\
        & \cmark 
            & \makecell[l]{Visible in peripheral vision}
            & \makecell[l]{Easy to use} \\
        & ? 
            & \makecell[l]{Auditory Modes, icons}
            & \makecell[l]{-} \\
         & ? 
            & \makecell[l]{Heads-up Display}
            & \makecell[l]{Easy to read} \\
        & \cmark 
            & \makecell[l]{Use of colors$^\dag$ to indicate changes}
            & \makecell[l]{Easy to understand, Decreases mental load} \\
        & \cmark 
            & \makecell[l]{Comfortable$^\dag$ Acceptable Error bar}
            & \makecell[l]{Easy to use} \\
         \midrule
         \multicolumn{4}{c}{\cmark\quad Recommended \qquad\xmark\quad Not Recommended \quad \idkmark Controversial \quad\dag Personalized} \\
        \bottomrule
         \\
    \end{tabular}
     }
\end{table*}
All policies were trained in simulation and evaluated in both simulation and through a driving simulator user study and are shown to be provably effective in mitigating congestion with no sim-to-real transfer\footnotemark[1].
Particularly, the policies are built on the FLOW framework~\cite{flow} that allows for simulation with the SUMO traffic simulator~\cite{sumo} and co-simulation~\cite{sumo_cosim, hcd-workshop} with the CARLA Simulator~\cite{carla}. 


}

\end{document}